\documentclass{aastex6}
\usepackage{graphicx}
\usepackage{amsfonts}
\usepackage{natbib}

\newcommand{\etal}{et al.}
\newcommand{\gal}{$\alpha$}
\newcommand{\gb}{$\beta$}
\newcommand{\gla}{$\lambda$}

\shorttitle{CHROMOSPHERES AND OXYGEN}
\shortauthors{Dupree, Avrett, Kurucz}

\begin{document}
\slugcomment{ApJ Letters in press}
\title{Chromospheric Models and the Oxygen Abundance in Giant Stars}

\author{A. K. Dupree, E. H. Avrett, and R. L. Kurucz }
\affil{Harvard-Smithsonian Center for Astrophysics, Cambridge, MA
  02138, USA}
\email{dupree@cfa.harvard.edu}

\begin{abstract}
Realistic stellar atmospheric models of two typical metal-poor giant stars 
in Omega Centauri that 
include a chromosphere (CHR) influence the formation of  optical  lines of
\ion{O}{1}: the forbidden lines ($\lambda$6300, $\lambda$6363)   and
the infrared triplet (\gla\gla7771$-$7775). One-dimensional 
semi-empirical non-LTE models are constructed
based on observed Balmer lines.   A full non-LTE formulation is 
applied in evaluating line
strengths of O I  including photoionization by the Lyman continuum and photoexcitation
by Ly-\gal\ and Ly-$\beta$.   Chromospheric models (CHR)
yield forbidden oxygen 
transitions  that are  stronger than
in radiative/convective equilibrium (RCE) models. The triplet 
oxygen lines from high levels  also appear stronger 
than produced in  an   
RCE model.   The inferred oxygen abundance from realistic CHR models for these two stars
is  decreased by factors $\sim$3 as compared to values 
derived from RCE models. A lower oxygen abundance suggests that intermediate mass AGB stars
contribute to the observed abundance pattern in globular clusters.  A change in the 
oxygen abundance of metal-poor field giants could affect models of deep mixing episodes 
on the red giant branch. Changes in the oxygen abundance can impact other abundance determinations 
critical to astrophysics including chemical
tagging techniques and galactic chemical evolution. 
 
\end{abstract}


\section{Introduction}
Determination of stellar abundances constitutes a fundamental component
of a wide variety of issues in  contemporary astrophysics. 
Large spectroscopic surveys of chemical  compositions can reveal
the history of star formation in our Galaxy and the early 
universe (Frebel \& Norris 2015).  
Ages of clusters are inferred from chemical abundances ( Fulbright \& Johnson 2003; 
Maderak et al. 2015; Marino et al. 2012a, 2012b).  The origin of multiple 
populations in globular clusters is addressed by abundances and 
remains not understood (Gratton et al. 2012, Renzini et al. 2015).

The \ion{O}{1} atom  produces two sets of transitions in stars that are 
frequently used to evaluate the oxygen abundance: the forbidden [O I] 
lines at 6300\AA\ and 6363\AA\  and the  \ion{O}{1} triplet at 7771$-$7775\AA.  
Abundances inferred  using radiative/convective equilibrium (RCE) models 
yield  values  in giant stars that 
can differ by 0.3 to 1.2 dex depending on the oxygen transition  
(Fulbright \& Johnson 2003; Takeda 2003; Israelian et al. 2004).  
While it has been  asserted that the  
forbidden transitions appear to be in local
thermodynamic equilibrium (LTE)  using photospheric models 
(Kiselman 2001; Takeda 2003; Asplund 2005), 
non-LTE effects can  create an apparently
stronger triplet transition in a RCE model (Takeda 2003, 
Shchukina et al. 2005; Schuler et al. 2006;
Fabbian et al. 2009; Amarsi et al. 2015) requiring a slightly lower oxygen abundance. 
Takeda (2003) noted that  even taking into
account  non-LTE corrections in a RCE model,  
metal-poor disk/halo stars  still exhibit
systematic anomalies between diagnostics that can amount to a median value of
$\sim$0.3 dex.
Authors have conjectured that these
discrepancies arise from inadequate  model atmospheres or in the
calculation of spectral lines
(Israelian et al. 2004; Shchukina et al. 2005;  
Schuler et al. 2006).

Critical to the determination of  abundances are the 
stellar models used to extract abundance values. Stellar models 
(i.e. MARCS, PHOENIX, ATLAS)
traditionally consist of a one-dimensional  radiative and
convective equilibrium (RCE)  stellar photosphere in hydrostatic
equilibrium. The temperature decreases from the interior of the star 
outwards with constant radiative and convective flux and with gravity balancing the
pressure gradient (Gustafsson et al. 2008; Castelli \& Kurucz 2004; 
Hauschildt et al. 1999; M\'esz\'aros et al. 2012).
The level populations and line strengths are computed in LTE.  
In some instances, adjustments are
made to construct so-called non-LTE corrections 
to the RCE models for \ion{O}{1} (Takeda 1997; Gratton et al. 1999), 
usually with simplifying assumptions.

Another development includes the construction of three-dimensional models of
a stellar photosphere with hydrodynamical simulations of radiative and
convective transport  (Asplund et al. 1999;  Freytag et al. 2010).  
Such models   have been used (cf. Dobrovolskas et al. 2014) to  
assess the impact of inhomogeneities in the
photospheres in cool  stars and the appropriate correction factors for abundances
when derived from a 1-D model. 

{\it However, these previous models for giant stars are unrealistic because 
chromospheres are missing.}   Many if not all cool stars 
exhibit spectroscopic signs of  non-radiative heating which
creates chromospheres with temperatures in excess of  the
photospheric temperature.  Optical stellar spectra signal the presence of chromospheres 
in a  variety of ways:  emission in the
cores of the \ion{Ca}{2} H and K lines and \ion{Na}{1}; 
emission in the wings of H$\alpha$; and the presence of near-ir \ion{He}{1} 
(Dupree et al. 1984; Dupree et al. 2007; Dupree \& Smith 1995; 
M\'eszaros  et al. 2009; Mart\'inez-Arn\'aiz et al. 2010).  

In this paper, we make the first 
calculations of optical line
profiles for oxygen in two giant stars using  a realistic  atmospheric model 
that includes a chromosphere (CHR). 
We chose the O I atom because optical 
transitions in oxygen are widely used in stellar  abundance determinations of cool 
luminous stars (Boesgaard et al. 2015, Korotin et al. 2014, VandenBerg et al. 2014; 
Johnson et al. 2013; Johnson \& Pilachowski 2010, 2012; Gratton et al. 2015).
Oxygen, the third most abundant element, remains pivotal in the
astronomical context. Some have suggested that
oxygen is superior to iron as a tracer of galactic chemical evolution because
the sites for its synthesis are better understood (Wheeler et al. 1989).

\section{The Calculation Strategy}

Oxygen level populations are calculated assuming a 
radiative/convective equilibrium (RCE)
model in LTE to provide comparison profiles. Then 
an atmospheric model is constructed with a chromosphere (CHR)     
to match the observed H\gal\ and H$\beta$  line profiles of two red giants
in Omega Centauri. Because these Balmer lines are produced in a wide 
region of the chromosphere,  they link the photospheric model
to that of the chromosphere and define a reliable semi-empirical model. 
This non-LTE model based on hydrogen is  used to calculate
the  oxygen populations and the line profiles, treating  oxygen
as a trace element.   The effects of the Lyman continuum and
the Ly-\gal\ and  Ly-$\beta$ line (Bowen fluorescence), calculated in full non-LTE
approximation, are 
included in the ionization and excitation of oxygen.

\section{The PANDORA Code}
We use the PANDORA code (Avrett \& Loeser 2008) to construct
a one-dimensional semi-empirical atmosphere  by iteratively evaluating for the 
hydrogen atom, the ionization
state, line and continuum radiation, and  level 
populations. 
The atmospheric structure ({\it T}, {\it N}$_{\rm H}$, and 
turbulent broadening)  is modified until the calculated   
line profiles agree with observation. The value of {\it N}$_{\rm H}$ is determined empirically
rather than by hydrostatic equilibrium.   The total model is iterated
with full non-LTE calculations in order to
match the specified line profiles. All transitions are calculated
explicitly.  Calculations are carried out in plane-parallel geometry, evaluating
the profile at various values of $\mu = cos\ \theta$, and then integrating over
all angles.

\section{The Atomic Models and  Rates}

The atomic model for \ion{H}{1} replicates that used in Avrett \& Loeser (2008)
with a 15-level H atom plus continuum that involves 105 line 
transitions. Partial frequency redistribution is used for the calculation 
of the Ly-\gal\ and Ly-\gb\ profiles. The Lyman continuum is
also included as it affects the hydrogen ionization.  
A calculation for hydrogen is made first  in order
to determine the ionization structure in the atmosphere by iterating to fit the
observed H\gal\ and H$\beta$ profiles. The Lyman series profiles are evaluated
explicitly.  Once the ionization
structure and the electron density are known, then  level populations and
line profiles for trace atoms such as O I, can be calculated. 

The atomic model for \ion{O}{1} consists of   40 individual energy levels grouped
into  20 reference levels.  This includes the ground level of \ion{O}{2}
and two levels in the continuum of \ion{O}{1}.  Electron and hydrogen collisions between all levels
are included, as is collisional ionization by electrons and 
hydrogen and photoionization from all levels including effects of the Lyman continuum. 
Radiative and dielectronic  recombination are included.
Electron-impact excitation rates are taken from calculations of Barklem (2007),
and supplemented by those of Seaton (1962). Hydrogen-impact rates are from Kaulakys (1985) 
and Drawin (1969). Dielectronic recombination rates are taken from Badnell et al. 
(2003) and their fitting parameters were employed.

Of particular import for neutral oxygen is the  Bowen mechanism (Bowen 1947)
in which the coincidence in wavelength between Ly-\gb\ and the \ion{O}{1} 
transition from the ground state $^3$P to an excited 
$^3$D level ({\it 2p$^4$ $^3$P - 2p$^3$3d $^3$D}) increases the 
populations in the triplet sequence which also contributes to the intersystem 
collision rates.  The strength
of the \ion{O}{1} emission in the first ultraviolet spectrum of a giant
star, \gal\ Boo, was found to be surprisingly strong (Moos \& Rottman 1972) and 
was subsequently understood to result from Ly-$\beta$ pumping as
proposed by Bowen (Haisch et al. 1977).   
UV surveys document the dominating
strength of the \ion{O}{1} resonance lines in cool luminous stars (Ayres et al. 1995) and 
require inclusion of photoexcitation and photoionization from chromospheric hydrogen radiation 
in the calculation of oxygen level populations and ionization.

\section{Atmospheric Model}
A model was constructed loosely based on two giant stars in the globular
cluster Omega Centauri. We select a model with {\it T}$_{eff}$=4745K, 
{\it log g} = 1.74, [Fe/H]=$-$1.7, and $\alpha$-enhancement 
([$\alpha$/Fe]= +0.4). 
These parameters are consistent with spectroscopic analyses of LEID 54084 
and LEID 54064 in Omega Centauri (Johnson \& Pilachowski 2010; Dupree et al. 2011).
Solar abundances
(Asplund et al. 2009) are 
decreased by $-$1.7 dex and the alpha elements (Ne, Mg, Si, S, Ar, Ca, and Ti) 
are enhanced by +0.4 dex relative to the scaled solar values. 
The abundance of oxygen is first taken as 
log O=7.64 where log H=12.00; subsequently this abundance is changed in 
additional calculations.  
A radiative/convective equilibrium (RCE)  and a chromospheric model (CHR) 
were constructed.
The temperature distributions for these models are shown in Figure 1 and discussed below.

\subsection{Radiative/Convective Equilibrium Model (RCE)}
This  model  in hydrostatic equilibrium    
exhibits a monotonically decreasing temperature with height.
The Kurucz model (Castelli and Kurucz 2004) was  calculated specifically
for the stellar parameters listed above. 
Values of turbulent broadening are introduced
to match  the observed width of the Balmer lines in the stars. 
The RCE model  does not produce good fits to  the
Balmer line profiles. It has long been known (Dupree et al. 1984) that the
RCE approximation is not expected to
match the observed profiles because  Balmer lines have a 
chromospheric component.

\subsection{Chromospheric Models (CHR)}

The construction of a semi-empirical chromospheric  model (CHR) relies on 
matching appropriate line profiles.  
The H\gal\ and H$\beta$ lines are useful in this regard since the atmospheric levels of 
profile formation span
a large temperature region in low gravity stars (M\'eszaros et al. 2009). The RCE model 
is extended in height to 1.8R$_\star$ (where R$_\star$=30R$_\odot$) along with a 
temperature increasing to 10$^5$K 
with height.  We solve  the non-LTE optically thick transfer equation for hydrogen
and modify the atmospheric model (Fig. 1) to produce good agreement with 
the H\gal\ and H$\beta$  line profiles (Fig. 2).  These profiles 
originate from the high photosphere in the wings of
the line, traverse
the temperature minimum, and arise from the low chromosphere in the line
core as shown in Fig. 1. We take this semi-empirical model as the 
basis for the oxygen calculations.
The Ly-$\beta$ profile from the hydrogen model  is used as the input for 
photoexcitation of the
$^3$D level of \ion{O}{1}.   The Lyman continuum and Ly-\gal\ from the hydrogen
model are also included to evaluate the oxygen ionization and excitation.

A recent
calculation of oxygen in metal-poor stars using a 3D RCE 
model (Amarsi et al. 2016) notes the 
importance of the Ly-\gal\ wings in photoexcitation of the oxygen resonance lines.
A RCE model, lacking a chromosphere will not produce an accurate Lyman continuum value nor
a realistic Ly-\gal\ profile,
so its influence is difficult to assess. In addition,  the Ly-\gal\ profile
needs to be calculated with partial frequency distribution which will lower the
flux in the line wings. Unfortunately, the Ly-\gal\ profile is challenging to measure 
directly because of interstellar absorption and geocoronal emission.

\section{Oxygen Line Profiles}

Line profiles for three oxygen transitions are shown in Figure 3.
The CHR solutions are stronger than
the RCE/LTE profiles for all transitions. 
The forbidden transitions  reach optical depth unity  deep in the atmosphere 
where higher densities and greater collisional coupling occurs.   
However, the levels are not in their LTE population ratios where the forbidden lines
are formed (Fig. 1). The strengthening of  the triplet transition has been noted
previously in RCE models by a number of authors (Takeda 2003, Fabbian et al. 2009), but
the chromospheric model strengthens the lines to a greater degree. 

To evaluate the sensitivity of the line strength to the CHR model, we 
changed the temperature  over the region of
formation of the oxygen lines in steps of 100K extending $\pm$500K from the
model.  In order to be acceptable, the H\gal\ and H$\beta$ 
lines are well replicated (their equivalent widths changed by only $\sim$2\%).
The equivalent widths of the oxygen lines changed by an average of 9\% for
these models  suggesting that the strength of the oxygen lines depends 
principally on the abundance of oxygen. If
Ly-$\beta$ pumping is omitted, the equivalent widths of the triplet 
decrease by $\sim$ 20\%.   In conclusion, the model is reasonably 
fixed by the hydrogen lines and changing the 
temperature where the oxygen lines are formed creates a negligible difference
in the equivalent widths of the oxygen lines. 

The calculated profiles for 3 values of the O abundance (Fig. 4) are 
compared to observations of the Omega Cen 
giants, LEID 54064 and LEID 54084. Equivalent widths from the models are
given in Table 1.  The spectra 
from Magellan/MIKE have S/N reaching 100-200, but the
oxygen lines are weak, and the forbidden lines are challenged
by blending with other atomic lines, telluric lines (\gla6300), and a continuum
marred by a broad autoionizing line of \ion{Ca}{1} (\gla6362). However, the 
observations of both forbidden transitions and the triplet  indicate 
generally consistent agreement with model profiles having similar values of the
oxygen abundance.     Synthesis of the \gla6300 transition (Johnson and Pilachowski 2010)  
for LEID 54064 and LEID 54084 suggests log O/H=7.53 and 7.63 respectively where log H=12.00.  
From our CHR model, the \gla7775 multiplet yields  log O/H = 7.04 in LEID 54064, a reduction
of $\sim$0.5 dex; the forbidden transitions and the \gla7775 multiplet in 
LEID 54084 also appear
consistent with a low abundance: $\sim$7.1, a reduction also by $\sim$0.5 dex.
These CHR models  eliminate the abundance differences found (Takeda 2003) 
between forbidden and triplet lines using  RCE/NLTE  models. The abundance changes 
suggested here by the 1D CHR/NLTE model are 
larger for both the forbidden and triplet transitions than
those deduced from 3D/NLTE photospheric models of a subgiant star  
(Amarsi et al. 2015) which
indicate essentially no change in the abundance from the 
forbidden lines when compared to 1D/LTE models and a more modest correction 
($\sim -$0.14 dex) for the triplet.

\section{Discussion and Conclusions}

Adoption  of a stellar atmospheric model with a  chromosphere, 
in  a  1D non-LTE calculation 
of level populations and line strengths for \ion{O}{1}, including 
photoexcitation by  Ly$\beta$ 
affects both the forbidden transitions (6300\AA\ and 6363\AA) as well as the
triplet transition (7771--7775\AA).  These transitions are  systematically changed  
as compared to a RCE model: a 
chromosphere strengthens both the forbidden transitions and  the
triplet transitions. With the (slightly) increased temperature in 
the CHR model, higher
levels of \ion{O}{1} are populated. 

These results challenge the often cited statement (Kiselman 2001) 
that the forbidden lines ($\lambda$6300 
and $\lambda$6363) can be analyzed with an LTE approximation.  While this may
be true for a photospheric model, the presence of a chromosphere 
increases the forbidden lines substantially.     
Thus ``LTE abundances'' will overestimate the true oxygen abundance by a factor
of $\sim$3 in these stars.

Our results suggest that the inferred oxygen abundance is decreased when a 
CHR model is invoked.  Since differing H-alpha profiles and variation in 
H-alpha profiles in red giants suggest  intrinsic changes in the dynamics, 
and most likely the temperature and density structure of the stellar 
atmosphere, it is reasonable to expect that such changes signal different 
chromospheric structures and  the strengths of the oxygen lines may also
differ between similar metal-poor red giants. Of course there is no reason to 
believe that the absolute abundance of O changes due to chromospheric 
variability in a given star.   Chromospheres in metal-poor field stars are 
anomalously `strong' considering the lower metal abundance of these objects, 
and the chromospheric radiative losses do not appear to scale with 
metallicity (Dupree et al. 2007).  Since chromospheric structures can 
differ, we conjecture that the 'correction' for the oxygen abundance may 
not have a constant value, even for similar stars.  Thus one might expect 
that a Na/O anticorrelation may `spread out' 
due to chromospheric variation. Further calculations are necessary to
explore these effects.  The distribution could  become bimodal or
multimodal much as photometry reveals multiple populations in clusters.

Metal-poor giants in the field do not display the typical Na-O anti-correlation 
found in globular cluster giants (Kraft 1994).  A possible decrease in the oxygen abundance
could suggest (Gratton et al. 2000) that deep mixing episodes occur on the upper
red giant branch to bring processed
material from the  CNO cycle to the surface; alternatively, abundance patterns
might be attibuted to the initial composition of the star-forming material 
from a previous generation of more massive stars.

The effects of non-LTE and chromospheres have been considered for dwarf stars of solar
metallicity.  Maderak et al. (2013) concludes that  oxygen abundances derived 
from the triplet using photospheric models, if corrected for non-LTE, may be 
reliable for solar-type dwarf stars with
solar metallicities.  Recent calculations
of the triplet for a solar model including  a chromosphere suggest less 
than 1.5\% difference in the equivalent widths for a non-LTE calculation 
(Dobrovolskas et al. 2014).  However, it is not stated whether
Lyman-$\beta$ pumping was included, and the case of low metallicities
remains to be considered.

Multiple populations in globular clusters and their origin 
are also assessed in part by  oxygen abundances.
A decrease in the O/H abundance of $\sim$0.5 dex, 
has the potential to affect
many abundance relationships (cf. Fig. 3, Gratton et al. 2012) and in 
particular the important Na/O anti-correlations. If all oxygen abundances in cluster  
giants were to be uniformly reduced by 0.5 dex, this would imply that the oxygen 
yield from Type II SN (Nomoto et al. 2006) is  overestimated and cannot satisfy 
the abundance patterns. An increase in the contribution of  AGB stars 
of $\sim$5M$_\odot$ would appear to be necessary (Johnson and Pilachowski 2010).
Our two target stars have similar oxygen abundances, but different values
of Na/Fe, thus challenging the Na-O anti-correlations.  Sodium is 
enhanced in the He-rich star (LEID 54084).  
Johnson and Pilachowski (2010) found  [Na/Fe]=+0.25 (LEID 54084) and 
[Na/Fe]=+0.16 (LEID 54064).  Another determination (Dupree et al. 2011) for the same 
stars, yielded [Na/Fe]=+0.37 (LEID 54084) and [Na/Fe]=$-$0.14 (LEID 54064).  
The sense of the variation agrees between both 
determinations, but the difference is larger from the later result. The typical
abundance pattern suggests that enhanced He accompanies enhanced Na, and
enhanced Na should correlate with a lower O abundance (Gratton et al. 2012)  
which it does not in these two target stars.
Abundance patterns in the two target stars are
consistent with respect to He and Na, but display similar lowered 
oxygen abundances.  However 
a note of caution emerges in that the radial distribution of the abundances in Omega
Centauri differs between oxygen and sodium (Johnson \& Pilachowski 2010)  
suggesting that a clean anti-correlation may actually not exist. 

Accurate abundance values 
may shed light on the discreteness of photometry of globular clusters
versus the continuity of abundance determinations (Milone et al. 2015; Renzini et al. 2015).
A wider range of CHR models can address the precise abundances  necessary
for understanding the hosts of exoplanets 
(Bedell et al. 2014; Teske et al. 2014).
Other elements merit  analysis with realistic chromospheric models too 
such as \ion{Fe}{1}, \ion{Fe}{2}, \ion{Na}{1}, and \ion{Mg}{1}. 
To follow up these results, a broader array of models must be evaluated.  At present,
we cannot assume the structure of a stellar chromosphere {\it a priori}, but 
it must be constrained by spectroscopic observations of the transitions 
typical of chromospheric 
temperatures: \ion{He}{1}, \ion{Ca}{2}, H\gal, Na D, \ion{Mg}{2} etc.   
A confounding parameter could be activity variations:  the chromosphere 
itself changes structure and the impact 
on the oxygen line strengths needs to be assessed. Thus differential abundances
could be suspect.   If a chromosphere results from magnetic
dynamo activity, one might suggest that older, metal-poor stars would experience
a decay in activity.  However, a temperature rise could result from pulsations
and shock formation. In luminous  metal-poor stars, the H\gal\ line 
emission varies in strength and asymmetry (cf. M\'esz\'aros et al. 2009), most likely from 
modest pulsations.  In addition, weak magnetic
fields detected in luminous stars (cf. Landstreet 2015) could be present in the oldest stars
and cause heating (Musielak et al. 2002). In any case, with a good understanding of the 
chromospheric structure, the oxygen line strengths and others can be 
calculated with confidence.

We appreciate the thoughtful comments of Christian Johnson on this research.

\facility{Magellan:Clay (MIKE)}


\begin{figure}
\begin{center}
\includegraphics[angle=0,scale=0.6]{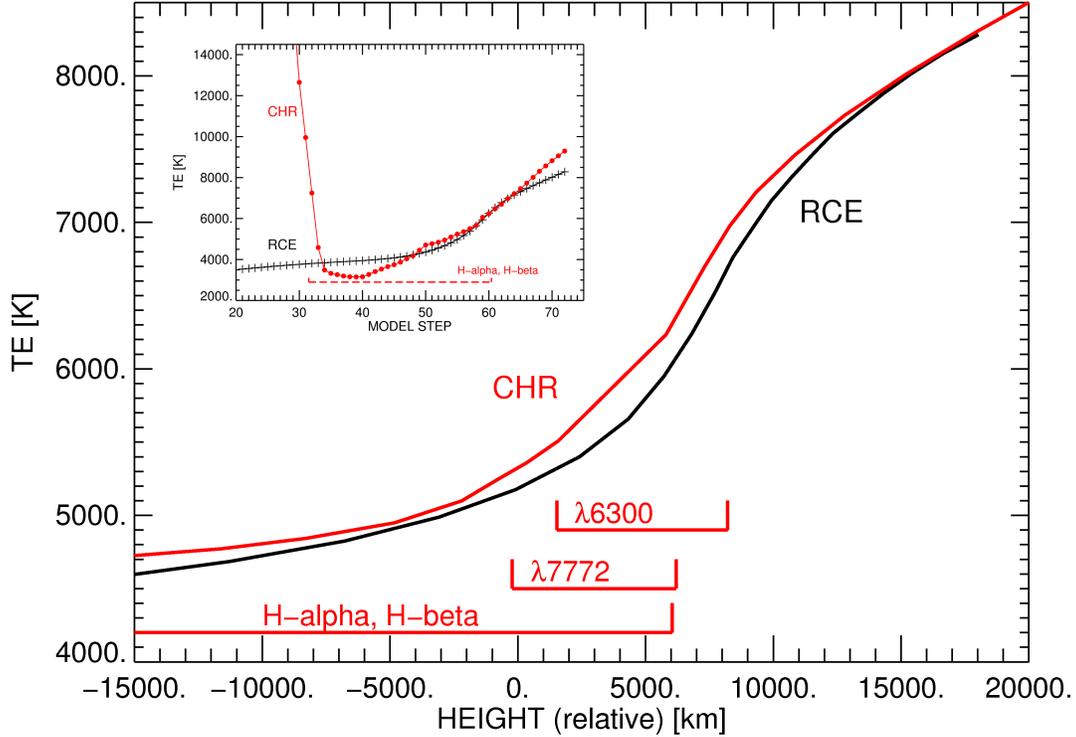}
\caption{Region of formation of the O I lines and H\gal\ and H$\beta$ in the CHR model. This region
is defined by the full width at half maximum  of each line contribution function for a disk-centered
profile taken at line center and $\pm$1\AA\ from line center. The RCE model
has been moved outward in height for illustration. In the CHR model, Height= 0 corresponds 
to $\tau_{5000}$=1. The CHR model extends much beyond the RCE model to $\sim$1.8 R$_\star$. }
\end{center}
\end{figure}


\begin{figure}
\begin{center}
\includegraphics[angle=0,scale=0.6]{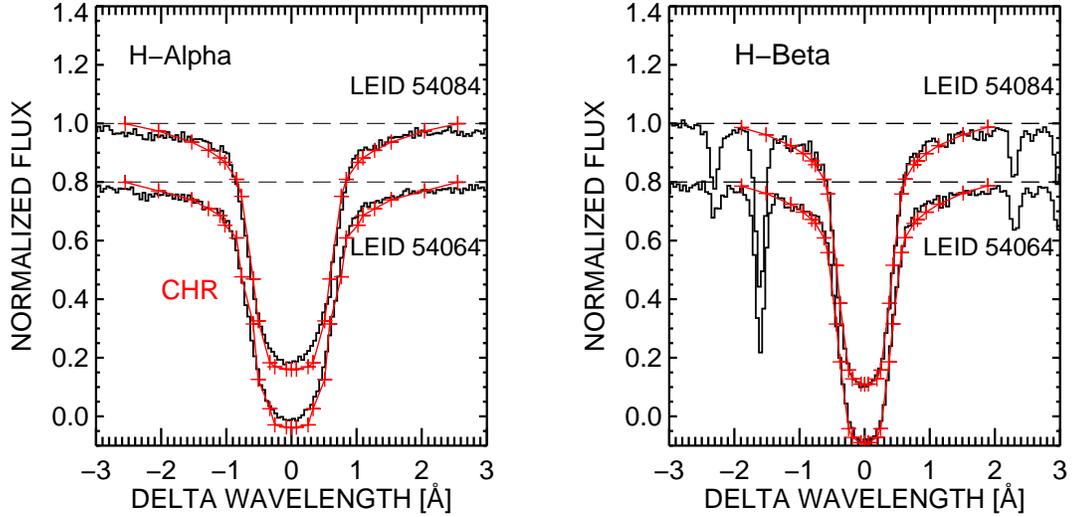}
\caption{H\gal\ and H$\beta$ profiles as observed (Dupree \etal\ 2011) in two Omega Cen giants, LEID 54084 and LEID 54064 ({\it black curves}), and the
calculated profiles using the same chromospheric model ({\it red curves}). The profiles for
LEID 54064 are offset by $-$0.2 for display.}
\end{center}
\end{figure}

\begin{figure}
\begin{center}
\includegraphics[angle=0,scale=0.7]{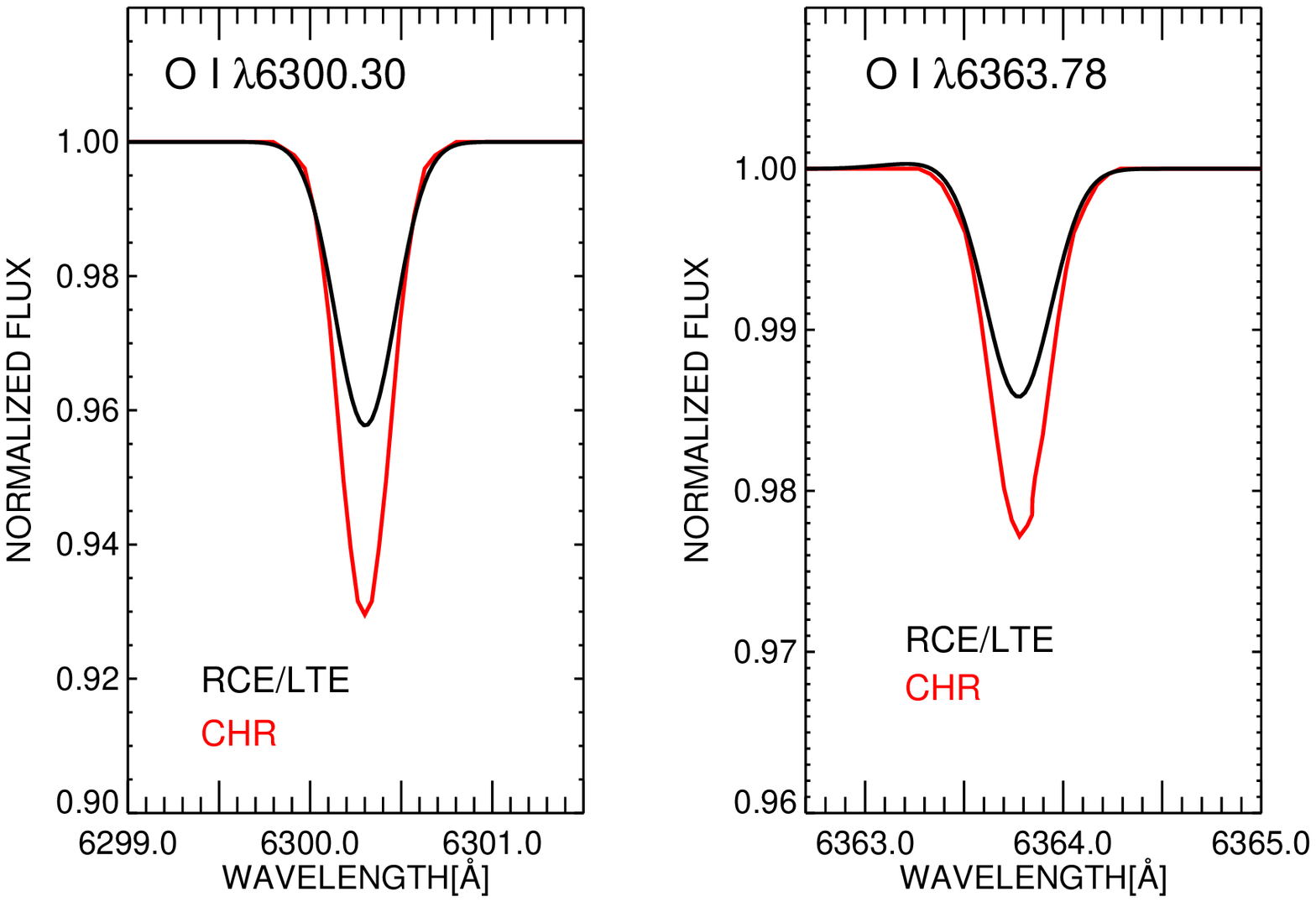}

\includegraphics[angle=0, scale=0.8]{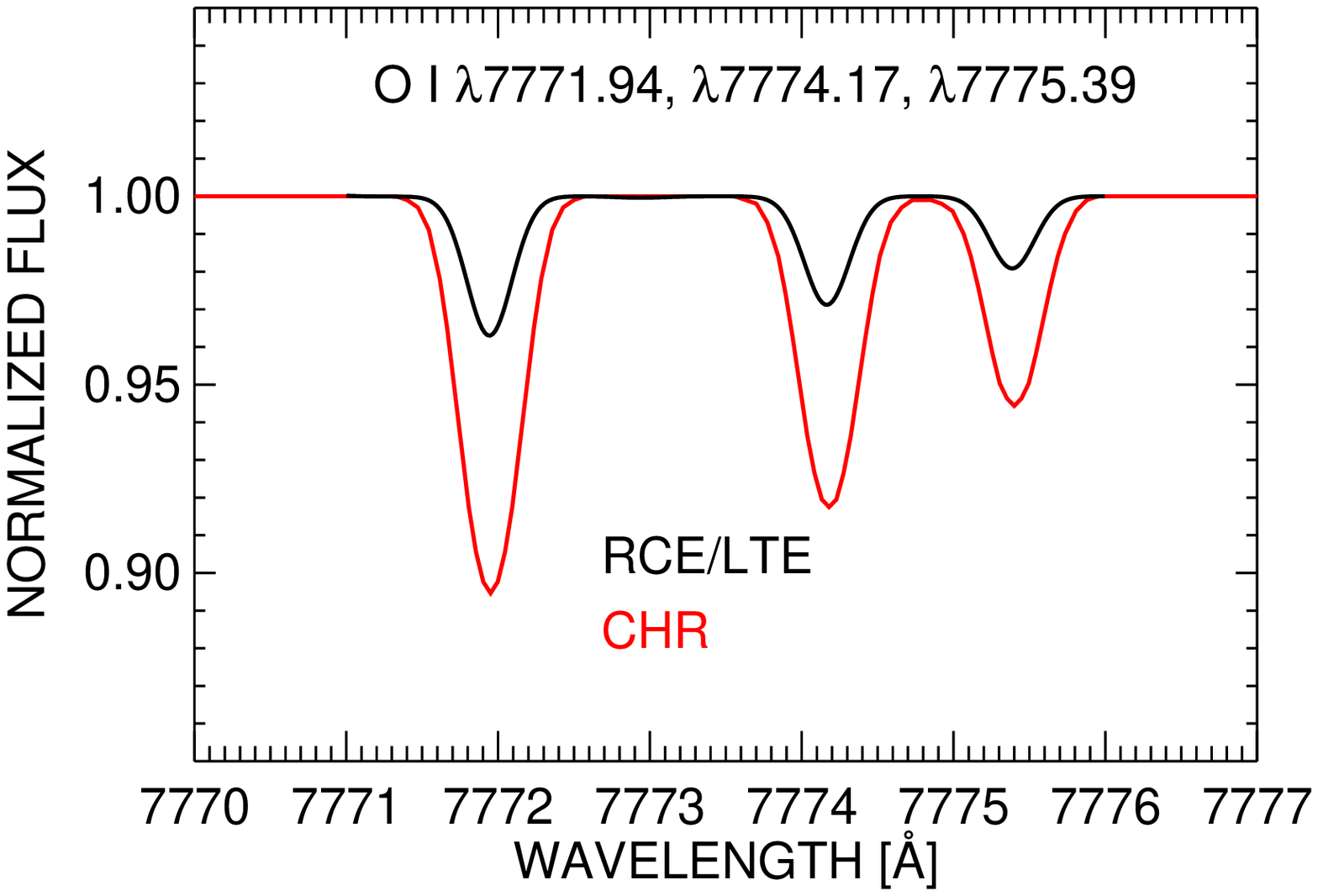}
\caption{Oxygen profiles for a RCE and CHR model. These are calculated for [Fe/H]=$-$1.7, and
are alpha enhanced; Oxygen abundance set to 7.64. All transitions are calculated (radiative). 
The CHR models produce stronger lines than RCE models in all transitions. }
\end{center}
\end{figure}

\begin{figure}
\begin{center}
\includegraphics[angle=0, scale=0.7]{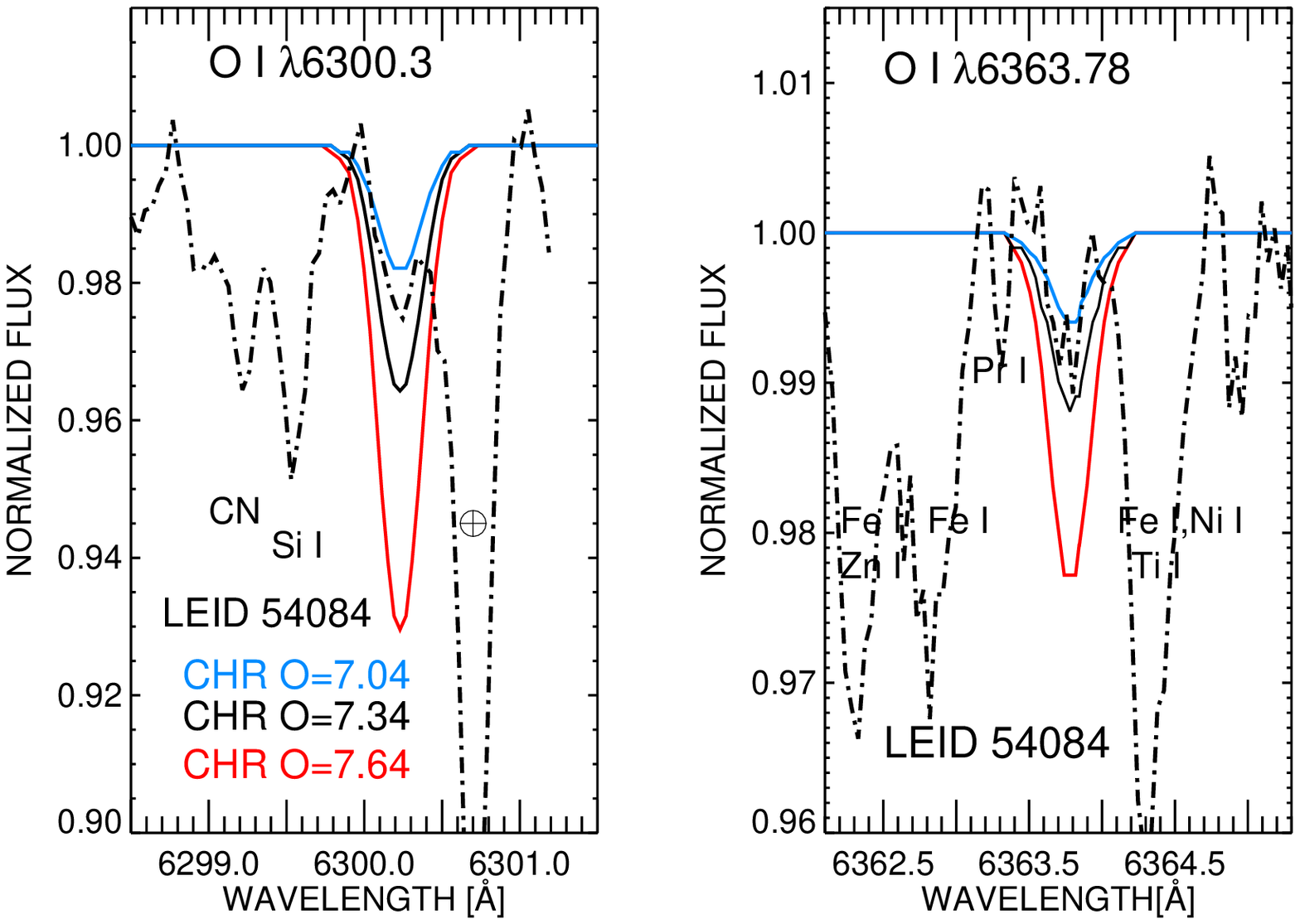}
\includegraphics[angle=0,scale=0.8]{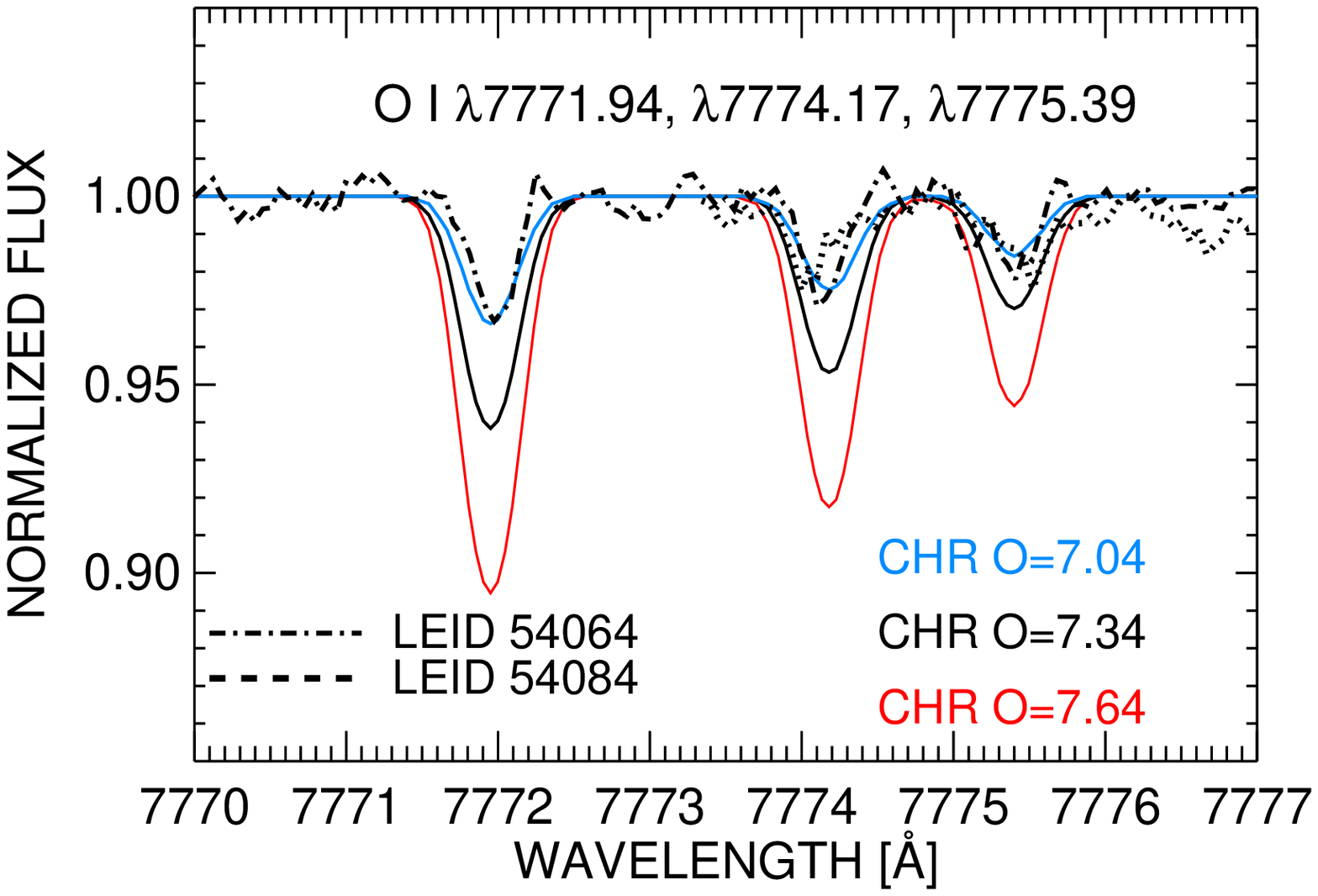}
\caption{Calculated oxygen profiles with a CHR model for various values of the oxygen abundance,
log O/H on a scale where log H=12.0.  
Observed profiles with Magellan/MIKE (Dupree et al. 2011)  for the target stars, LEID 54064 and 
LEID 54084 are shown.}
\end{center}
\end{figure}

\newpage

\begin{deluxetable}{llccccc}
\def\a{\phantom{0}}
\def\b{\phantom{00}}
\tablecolumns{7}
\tablewidth{0pt}
\tablenum{1}
\tablecaption{Equivalent Widths (m\AA) and Abundances from Atmospheric Models}
\tablehead{
\colhead{Model}&
\colhead{(O/H)}&
\colhead{\gla6300}  &     
\colhead{\gla6363}  &
\colhead{\gla7772}&
\colhead{\gla7774}&
\colhead{\gla7775}\\
\colhead{E.P.(eV)}& Model
& 0.0 &0.019&9.15&9.15&9.15\\
\colhead{log gf}&
&$-$9.78&$-$10.26&0.36&0.21&$-$0.02}

\startdata
RCE&7.64& 17.5 & 5.7&13.5&10.5&7.0 \\
CHR & 7.64 &28.7 &10.7&57.9&42.9&27.7 \\
CHR & 7.34 &14.2 &5.3& 31.1& 23.2&14.5 \\
CHR & 7.04 &7.4& 2.5&16.8&12.3&7.4\\
\sidehead{O/H Abundance}
RCE\tablenotemark{1}(LEID 54064)&&7.53&\nodata&\nodata&\nodata&\nodata \\
RCE\tablenotemark{1}(LEID 54084)&&7.63&\nodata&\nodata&\nodata&\nodata \\
CHR\tablenotemark{2}(LEID 54064)&&\nodata&\nodata&7.04&7.04&7.04\\
CHR\tablenotemark{2}(LEID 54084)&&7.1&7.1&\nodata&7.04&7.04
\enddata
\tablenotetext{1}{Johnson \& Pilachowski 2010}
\tablenotetext{2} {This paper.}
\end{deluxetable}
\end{document}